**Polar and phase domain walls with conducting interfacial states in a Weyl semimetal MoTe$_2$**

Fei-Ting Huang[1], Seong Joon Lim[1], Sobhit Singh[2], Jinwoong Kim[2], Lunyong Zhang[3], Jae-Wook Kim[1], Ming-Wen Chu[4], Karin M. Rabe[2], David Vanderbilt[2] and Sang-Wook Cheong[1*]

[1]Rutgers Center for Emergent Materials and Department of Physics and Astronomy, Rutgers University, Piscataway, New Jersey 08854, USA

[2] Department of Physics and Astronomy, Rutgers University, Piscataway, New Jersey 08854, USA,

[3]Laboratory for Pohang Emergent Materials and Max Plank POSTECH Center for Complex Phase Materials, Pohang University of Science and Technology, Pohang 37673, Republic of Korea

[4]Center for Condensed Matter Sciences, National Taiwan University, Taipei 106, Taiwan

Correspondence and requested for materials should be addressed to S.-W. Cheong (email: sangc@physics.rutgers.edu)

**Abstract**

**Much of the dramatic growth in research on topological materials has focused on topologically protected surface states. While the domain walls of topological materials such as Weyl semimetals with broken inversion or time-reversal symmetry can provide a hunting ground for exploring topological interfacial states, such investigations have received little attention to date. Here, utilizing in-situ cryogenic transmission electron microscopy combined with first-principles calculations, we discover intriguing domain-wall structures in MoTe$_2$, both between polar variants of the low-temperature($T$) Weyl phase, and between this and the high-$T$ high-order topological phase. We demonstrate how polar domain walls can be manipulated with electron beams and show that phase domain walls tend to form superlattice-like structures along the $c$ axis. Scanning tunneling microscopy indicates a possible signature of a conducting hinge state at phase domain walls. Our results open avenues for investigating topological interfacial states and unveiling multifunctional aspects of domain walls in topological materials.**

In the past decade, an explosion of research has focused on a sweeping search of candidate materials that may harbor topologically protected surface states[1-6]. The appearance of massless quasiparticles near topologically protected surface states are their key features, which could be two-dimensional (2D) Dirac points on the surfaces of topological insulators (TIs), or Fermi-arc surface states attached to the bulk Weyl points in the case of three-dimensional topological Weyl semimetals (WSMs)[3-5]. The manipulation of these surface states through homo/hetero-structures between topological phases promises functionalities going beyond those of their constituents with important applications such as dissipationless electronics[7-10]. For example, when these topological insulators are interfaced with superconductors, the emergent zero-energy Majorana fermions at the boundaries can be utilized for topological quantum computation[7]. The Veselago lens, which is the electronic lenses going beyond the diffraction limit, could also be realized through Weyl semimetal $p$-$n$ junctions[9]. Despite the concept of topological protection, to utilize topological surface states remains challenging due to the chemical/structural/electronic complexity of the surfaces[11,12]. Alternatively, domain walls of topological materials are self-assembled vacuum-free interfaces which can, in principle, replicate or facilitate new topological interfacial/edge states, but limited work has been done to date[13-15].

Among those topological materials, WSMs can be generated quite systematically in semimetallic crystals with the large spin-orbit coupling by breaking either time-reversal or space-inversion symmetry[3,5]. A considerable number of WSMs with broken inversion symmetry have been theoretically and experimentally identified, including transition-metal dichalcogenide (TMD) orthorhombic (Mo,W)(Te,P)$_2$ [16-18], transition-metal monopnictide (Ta, Nb)(As, P) family[4,6,19], and the RAlGe (R = rare earth) family[20,21]. An appealing aspect of these WSMs is that they also crystallize in polar crystallographic structures with a unique polar axis along which the two opposite directions are distinguishable, and thus they are polar WSMs. Note that since they are highly conducting at low frequencies, these polar WSMs belong to the so-called polar metals that have recently drawn much attention in the ferroelectric community[22-25]. In principle, itinerant electron screening in a (semi)metal might rule out the necessity of electrostatically-driven domain formation due to the fundamental incompatibility of polarity and metallicity, but the existence of polar domains, formed by local bonding preferences, is still possible since this mechanism is insensitive to the presence of charge carriers[22]. Some progress has been made in, for example, the polar interlocked ferroelastic domains observed in polar metal Ca$_3$Ru$_2$O$_7$ [25,26] and the structural defect-mediated polar domains in metallic GeTe.[27] In this context, exploring the domain structures in polar Weyl semimetal would be particularly important because the Weyl points and Fermi-arc connectivity can be manipulated via domain reorientation or locally modified order parameters at these DWs[28-31]. Solving the Weyl equation under the experimentally known DW geometry is highly desired.

Here we choose TMD MoTe$_2$, which has recently drawn immense attention due to its phase tunability and unique physical properties, such as extremely large magnetoresistance[32], superconductivity[33,34], high-order topology[35-37], the novel type-II WSM phase[16,17] and the polar metal (Supplementary Note 1 and Supplementary Fig. 1)[31,38,39]. Utilizing in-situ cryogenic transmission electron microscopy (TEM) and low-$T$ scanning tunneling microscope (STM), we unveil, for the first time, experimentally intriguing structures of polar domains and phase DWs between topologically distinct phases with metallic interfacial states in such a polar Weyl semimetal. We demonstrate the clear real-space ferroelectric reversible switching process controlled by the electron beam of TEM. The underlying physical mechanism is understood by combing first-principle calculations and group-theoretical analysis.

**Results**

**Unique layered structures of MoTe$_2$**

Depending on the crystal structure, MoTe$_2$ can be either in the semiconducting 2H or the semimetallic 1T' phase at room temperature. 1T'-MoTe$_2$ undergoes a first-order type structural transition from a monoclinic ($P2_1/m$, space group #11) to an orthorhombic polar T$_d$ ($Pmn2_1$, space group #31) structure at a critical temperature ($T_c$) of ~260 K (Supplementary Fig. 2). T$_d$-MoTe$_2$ is a rare simultaneous example of a material with superconductivity[33,34], a polar nature, and a topologically nontrivial band structure[16,17], whereas 1T'-MoTe$_2$ is a non-polar high-order topological material in which the 1D hinge instead of 2D surfaces host topologically protected conducting modes[35-37]. The type-II WSM transition occurs in the polar T$_d$ phase due to the requirement of broken inversion symmetry in this nonmagnetic system[16,40,41]. Notably, the inversion symmetry together with the time-reversal symmetry protects the higher-order topological phase in the nonpolar 1T'-MoTe$_2$[35-37]. Despite the apparent dissimilarity in the electronic structure, the 1T' and T$_d$-MoTe$_2$ phases can, in fact, be considered different stacks of the similar Te-Mo-Te layers. Figure 1a illustrates the basic unit, where the off-centered Mo atoms (blue spheres) move towards each other to form metallic zigzag chains (bold red lines) running along the *a* axis. Consequently, the Te octahedra are deformed with two possible orientations, denoted as P (**P**lus, arc counter-clockwise (CCW) arrow in orange) or M (**M**inus, arc clockwise (CW) arrow in purple), as shown in Fig. 1a.

Since P and M layers are translationally nonequivalent, MoTe$_2$ is a stack of P and M layers coupled through weak van der Waals forces with two possible interlayer shear displacements. Figure 1b presents the schematics of these shear displacements defined by the closest Te-Te ions (orange and purple dashed lines), denoted as + (positive; CCW displacement of Te-Te dashed line) or - (negative; CW displacement). 1T' and T$_d$ phases can be described by the stacking sequence counting from the bottom: two 1T' monoclinic twins are repeating arrangements of +M+P+ and -M-P- (1T'-I and 1T'-II in Fig. 1c), and two T$_d$-MoTe$_2$ polar states refer to stacks either as +M-P+ and -M+P- (T$_d$↑ and T$_d$↓ in Fig. 1c). Note that symbols P/M represent the intra-layer displacements of Te octahedra, which remain fixed, while +/- represent the variable interlayer shifts. A T$_d$ unit consists of +/- (i.e. different) displacements of the two sides of each layer (either P or M), while a 1T' unit has an identical interlayer displacement (either +/+ or -/- in Fig. 1c) and the preserved inversion center as marked. An uncompensated dipole, resulting in polarization along the *c* axis, can exist in the T$_d$↑ and T$_d$↓ states due to the asymmetric Te bonding environments triggered by the interlayer shifts (Supplementary Note 1 and Supplementary Fig. 1). This subtle difference of 1T' and T$_d$ phases has never been explicitly discussed and explains in part the reported sensitivity of the phases and electronic properties of MoTe$_2$ to external strain, pressure and thickness[32,33,42].

**Cross-sectional view of abundant phase domain walls**

Intriguingly, we find that 1T' twin walls have the T$_d$ character at room temperature. A cross-section view of 1T'-MoTe$_2$ has been imaged using dark-field transmission electron microscopy (DF-TEM) in combination with high-angle annular dark-field (HAADF) scanning transmission electron microscopy (STEM) imaging, which displays strong contrast associated with the atomic number of the local composition. Figures 2a-b reveal quasi-periodic monoclinic twin domains of alternating bright and dark bands along the *c* axis, which are consistent with the superposition of the diffraction spots (Fig. 2a) resulting from adjacent twin domains. 1T' twinning occurs by a mirror operation along the *ab*-plane. A further zoomed-in HAADF-STEM image of a twin wall (Fig. 2c) shows an atomically coherent interface between 1T'-I and 1T'-II along [110]. The interfaces (the blue lines in Fig 2c) can be readily identified by tracking the white-circled Mo positions. The yellow and orange shaded areas outline the 1T' monoclinic unit cells above and below the interfaces. It turns out that in addition to a mirror operation, a gliding of atomic layers is imposed on either side of the twin wall to reduce lattice strain. As a consequence, a thin planar $T_d$ unit (-M+) emerges owing to the crystallographic glide. The existence of a $T_d$ unit can also be understood in our notation in which the meeting region of 1T'-I (+ +, indicated by orange dashed lines between interlayers) and 1T'-II (- -, indicated by purple dashed lines) along the *c* axis naturally gives a layer with - + interlayer shearing (the schematic model in Fig. 2c). Thus, numerous $T_d$ mono-layer interfaces exist at the 1T' monoclinic twin walls at room temperature.

Next, we turn our attention to the temperature-driven 1T' to $T_d$ first-order phase transition, which manifests itself by resistivity anomalies with an evident thermal hysteresis[32,33,43] (Supplementary Fig. 2). A phase coexistence is expected within this hysteretic temperature window but remains little explored on meso- or nano-scales. Note that a pronounced thermopower enhancement near the phase boundary was ascribed to a significant gradient of scattering processes where real-space phase inhomogeneity may play an important role[44]. To explore the real-space phase configurations, we begin with cross-section views using in-situ cryogenic-TEM. Figures 2d-e are DF-TEM images taken in the same area at 80 K ($<< T_c$) and 300 K ($> T_c$) after a cooling/warming cycle. In the DF-TEM images using the stronger spots (green arrows in Fig. 2a), the areas associated with the major 1T'-I phase exhibit a bright contrast while regions with minor 1T'-II and newly nucleated $T_d$ phases remain dark upon cooling (from 300 K to 80 K). Interestingly, two essentially different types of periodicities consisting of alternating bright and dark stripes can be found. First, long-range stripes corresponding to two types of 1T' twin domains, exist at 300 K (Fig. 2b and Fig. 2e). At 80 K, additional short-range stripes appear inside individual twin domains (Fig. 2d). Instead of the $T_d$ phase growth from the existing $T_d$ units at twin walls, abundant thin-plate-like nucleation of the $T_d$ phase occurs within individual 1T' twin domains, in agreement with the appearance of additional diffraction spots (Fig. 3a).

Therefore, there appears an intimate connection between the 1T' and $T_d$ phases, and the system enters a metastable state with significant amounts of coexisting phase domains and DWs at the thermal 1T'-$T_d$ phase transition. The spatially modulating layers contain alternating $T_d$ and 1T' phases, resembling artificial thin-film superlattices. An enlarged 80 K DF-TEM image and the corresponding line profile are shown in Fig. 2f. The $T_d$/1T' phase modulation is rather periodic and the thinnest periodicity, 4-nm, consists of 6 layers of either the $T_d$ or 1T' unit. Possible atomic models are shown in Figs. 3b-3e. The experimental signature of the first-order phase transition manifests itself microscopically as a nanoscale modulation of in-phase (++/--) and anti-phase (+-/-+) interlayers with quasi-periodicity. Note that both 1T'-I and 1T'-II require a mechanical glide of the layers in opposite directions when transforming into the $T_d$ phase (Fig. 1c). The persistent phase coexistence at 80 K implies an effect of mechanical constraints applying restoring forces that tend to resist the layer-wise gliding from its initial position, particularly in our capped cross-section TEM specimen; details are given in Supplementary Note 2.

**In-plane view of polar domains and domain walls**

A further identification of polar states of these thin $T_d$ layers from cross-section views is beyond the detectability limits of our low-magnification DF-TEM technique. However, our *ab*-plane DF-TEM view reveals unambiguously the existence of two types of polar domains. Figure 4a displays the in-plane DF-TEM image of two domains with bright and dark contrasts, resulting from the non-equal diffraction intensity due to the broken space-inversion of $T_d$ phase at 80 K. Note that without initial cooling, no domains and DWs is found in any specimens at room temperature (Supplementary Figs. 3a-b). The domains with two different contrasts are associated with the ±*c* polar axes, but the absolute polarization direction cannot be identified in the *ab*-plane TEM view. Thus, for the sake of simplicity, we assign bright-contrast domains as $T_d$↑ and dark-contrast domain as $T_d$↓ in this work. The step-by-step phase transition during in-situ cooling is also provided in the sequential DF-images in Supplementary Figs. S4a-e. We also confirm the coexistence of 1T' and $T_d$ phase domains during a warming cycle (Supplementary Figs. 3b-d) in which the domain contrast of the 1T' phase remains intact in different imaging conditions. (More details are shown in Supplementary Fig. 3).

**Manipulation of polar domains and DWs with the electron beam**

The $T_d$ polar domains and DWs are found to be easily manipulated with in-situ e⁻ beam of the TEM at 80 K (Figs. 4a-b). The consistent and sharp domain contrast before (Fig. 4a) and after (Fig. 4b) an e⁻ beam irradiation suggests a still non-centrosymmetric structure, i.e. the $T_d$ phase. Figures 4c show four TEM snapshots from an in-situ video (Supplementary Movie 1) showing the shrinkage of a dark-contrast $T_d\downarrow$ domain, through a layer-by-layer gliding/phase-flipping process. A key feature is the observation of multi-DWs outlined by colored lines (Figs. 4c), which represent the boundaries between domains of different volume fractions of $T_d\uparrow/T_d\downarrow$ along the $c$ axis. Purple and blue shaded areas (Figs. 4c) mark $T_d\uparrow$ dominated and $T_d\downarrow$ dominated domains, respectively. The brighter contrast appears in DF-images when the $T_d\uparrow$ volume ratio is higher. The engineering of TMD polymorphs has attracted significant interest because of minimum-energy pathways or feasible transient polymorphs triggered by charge injection[45], laser irradiation[46], mechanical strain[47] and e⁻ beam irradiation[48]. In the case of MoTe₂, despite 2H to 1T' phase change can occur by laser irradiation due to a local heating and Te vacancies[46,47] or electrostatic gating[45], however, any transition mechanism involving the 2H phase is excluded in our work because of the consistent electron diffraction pattern and domain contrast before and after an e⁻ beam irradiation (Fig. 4a-4b and Supplementary Figs. 4f-4g). Notably, the induced domains and DWs return to their original morphology after spreading a focused beam. A restorative DW motion is captured by an in-situ video (Supplementary Movie 2). The reversibility of $T_d\uparrow$ and $T_d\downarrow$ domains proves that there is no massive Te atom loss or damage by the knock-on effect during the exposure. The e⁻ beam induced domain behavior is known in ferroelectric insulators, and attributed to positive specimen charging in insulating materials[49,50]; however, no static charge accumulation is expected in semimetallic MoTe₂.

To understand the switching phenomena, we compute the potential energy landscape of MoTe₂ using first-principles density functional theory (DFT) calculations[51-55] (see Methods). As shown in Fig. 1a, the P and M layers are related by the symmetry operation $M_z | \left(\frac{b}{2} + \lambda\right)$, where $M_z$, $b$ and $\lambda$ represent a vertical mirror, lattice vector, and the interlayer displacements between two neighboring layers, respectively. We identify a new high-symmetric orthorhombic structure $T_0$ (*Pnma*, space group #62) of MoTe₂ at $\lambda = 0$ (Fig. 1b), which belongs to the high-energy peak on the potential energy landscape of MoTe₂ (Fig. 4d). The $T_0$ phase has two instabilities: (1) an unstable in-plane optical phonon mode at the Brillouin zone center, and (2) an elastic instability yielding negative elastic stiffness coefficients (Supplementary Fig. 5). The first instability leads to an interlayer displacement of neighboring layers, yielding a double-well potential energy profile with two local minima at $\lambda = \pm 0.5$ Å, representing $T_d\uparrow/T_d\downarrow$ phases. The second instability causes a rigid shear of the orthorhombic unit cell, making the *b-c* cell angle non-orthogonal. By rotating the *b-c* angle of the $T_0$ phase, we again obtain a double-well potential energy profile having two local minima at 86.4º and 93.6º corresponding to 1T'-I and 1T'-II phases, and the predicted monoclinic angle is in good agreement with the experimental data (93.5º - 93.9º) [33,41,56]. This monoclinic distortion has the effect of shifting the neighboring layers horizontally, by about the same distance as in the $T_d$ phases, suggesting that it is driven by the same underlying microscopic instability.

Figure 4d shows the potential energy surface of MoTe$_2$ in the vicinity of the $T_0$ phase as a function of $\lambda$ and the *b-c* angle. We obtain four minima corresponding to $T_d\uparrow$, $T_d\downarrow$, 1T'-I, and 1T'-II phases, where the $T_d$ phases are the lowest in energy with reference to the high-energy point $T_0$. A direct structural transition from the $T_d\uparrow$ to $T_d\downarrow$ (1T'-I to 1T'-II) phase through the peak along path-1 (path-2), as shown in Fig. 4d, requires overcoming a large energy barrier of height 16.9 meV/u.c. (13.5 meV/u.c.). However, there are lower-energy pathways with an energy barrier of ~5 meV/u.c., marked as path-3 in Fig. 4e, suggesting that the $T_d\uparrow$ to $T_d\downarrow$ polar converting *via* an intermediate nonpolar 1T' phase is energetically preferable as shown in Fig. 4g. In this respect, a feasible low energy path through the 1T' DW-mediated switching process may be involved in our e$^-$ beam effect. The electron beam is certainly required to trigger the layer shearing.

We next consider the $T_0$, 1T' and $T_d$ phases from the view of symmetry. Figure 4f illustrates the MoTe$_2$ "family-tree" of the crystallographic group-subgroup relations[57]. The 1T' and $T_d$ phases reveal that a proper transition drives from the high-symmetric $T_0$ upon the $\Gamma_4^+$ or $\Gamma_2^-$ zone center instabilities, which is consistent with the phonon dispersion shown in Supplementary Fig. 5. A detailed symmetry analysis further indicates that *Pm* (space group #6) is a subgroup of both 1T' (*P2$_1$/m*) and $T_d$-MoTe$_2$ (*Pmn2$_1$*) and it is expected to link the 1T' and $T_d$ phases as shown in Fig. 4f. Space group *Pm* is, indeed, the symmetry to describe those superlattice-like structures appearing across transition (Fig. 3b), providing a complete unified symmetry description of MoTe$_2$.

**Phase domain wall conductance**

Finally, the atomic-scale electronic properties of lateral polar and phase DWs are also investigated by scanning tunneling microscopy (STM). In order to increase the density of polar and phase domains and DWs in the *ab*-plane, we have grown Fe-doped MoTe$_2$ crystals (see Methods and Supplementary Fig. 2). One polar/phase junction among T$_d$↑, T$_d$↓ and 1T' domains is found at 77 K (Fig. 5a) in a MoTe$_2$:Fe crystal with a slightly lower phase transition temperature (Supplementary Fig. 2). Consistent with the identical nature of each layer of 1T' and T$_d$ phases discussed above, three domains near the junction present similar topography and spectroscopic features as well as quasiparticle interference patterns, which are dominated by the atomic distribution of Fe dopants (the details are given in Supplementary Fig. 2 and Supplementary Fig. 6). On the other hand, DWs reveal two different types; the first type is marked with red and green dots in Figs. 5a-b), which deviates from the zigzag direction, i.e. the *a*-axis direction. The zigzags, corresponding to high-intensity lines in Fig. 5c, shift in the direction perpendicular to zigzags at the protruded area (red-dot DW) (Fig. 5c). The second type DW follows the zigzag direction (Figs. 5a-b blue). The shift at the first-type DWs is likely due to the mismatch of the unit cell between the monoclinic 1T' and orthorhombic T$_d$ phase domains. These considerations lead to the most likely domain assignment shown in Fig. 5a. From our TEM and STM observations, we find that polar DWs tend to be parallel along the zigzag direction while phase DWs tend to be highly curved (Fig. 4c and Supplementary Fig. 4).

Interestingly, from tunneling spectroscopy measurements, we observe characteristic local density of states at these two types of DWs, distinct from that of the bulk (Fig. 5d). The first-type protruded DWs, namely 1T'/T$_d$ phase DWs, (Fig. 5d red and green curves) show an enhanced conductance in the empty state while the second-type DW (Fig. 5d blue curve) does in the filled state. These features are also apparent in spatial mapping of conductance taken at filled (-100 mV, Fig. 5e) and empty (+100 mV, Fig. 5f) states. The systematic studies of spatial variation of local density of states are given in Fig. 6. Our results demonstrate distinct electronic properties at those polar and phase DWs in MoTe$_2$. Note that 1T'-MoTe$_2$ was earlier considered to be a topologically trivial material[16,17,41] based on the Fu-Kane $Z_2$ index criterion[58], however, recent theoretical works predict that 1T'-MoTe$_2$ inherits a higher order topological phase featuring topologically protected 1D hinge modes at the edges[35-37]. We notice that a considerably large conductance at the first-type 1T'/T$_d$ phase DWs with the orientation-dependent feature (Fig. 5d red and green curves). Alternatively, those protruded type 1T'/T$_d$ DWs can be promising candidates for the conducting hinge state studies in future[59].

**Discussion**

In summary, for the first time we report the existence of polar domains and abundant superlattice-like arrangements of phase DWs in MoTe$_2$ using in-situ cryogenic TEM along planar and cross-section views. We also discuss the feasible low-energy pathways of the polar domain switching. Our observations open up several important directions for future exploration. First, notably, the T$_d$ polar phases of MoTe$_2$ host topologically non-trivial Weyl points[16,17,40,41]. Since T$_d$↑ and T$_d$↓ polar phases are related by the space-inversion symmetry, Weyl points in these phases will have the same location in the energy and momentum space (but opposite chirality) and are hence considered "topologically identical." One naturally expects quantum phenomena occurring due to the projection of opposite pairs of Weyl points and the resulting Fermi arc patterns at the T$_d$↑/T$_d$↓ polar DWs. For example, as we tune this interlayer displacement parameter, λ, opposite Weyl points move towards each other in the momentum space, and finally mutually annihilate at λ = 0, i.e. no Weyl points for the T$_0$ phase. A similar manipulation of Weyl point separation and Weyl point number by interlayer displacements has been discussed in WTe$_2$[31]. In contrast to polar DWs, phase DWs are the interfaces between "topologically distinct" phases: topologically nontrivial WSM and high-order topological phases (i.e. T$_d$/1T' superlattice structures along the *c* axis). Our STM observations, which imply the possible presence of conducting hinge states in the 1T'/T$_d$ phase DWs, call for further attention. Second, those T$_d$/1T' superlattice regions with abundant phase DWs can be described as a transient state, which may be a rich area for macro-scale ordering by modulating the interlayer stacking and topological invariant. Lastly, the existence of polar domains and the electron beam manipulation of those polar DWs offer the possibility of rapid/controllable topological switching through electronic/optical excitations[31] and could be extended to other WSM or polar metals.

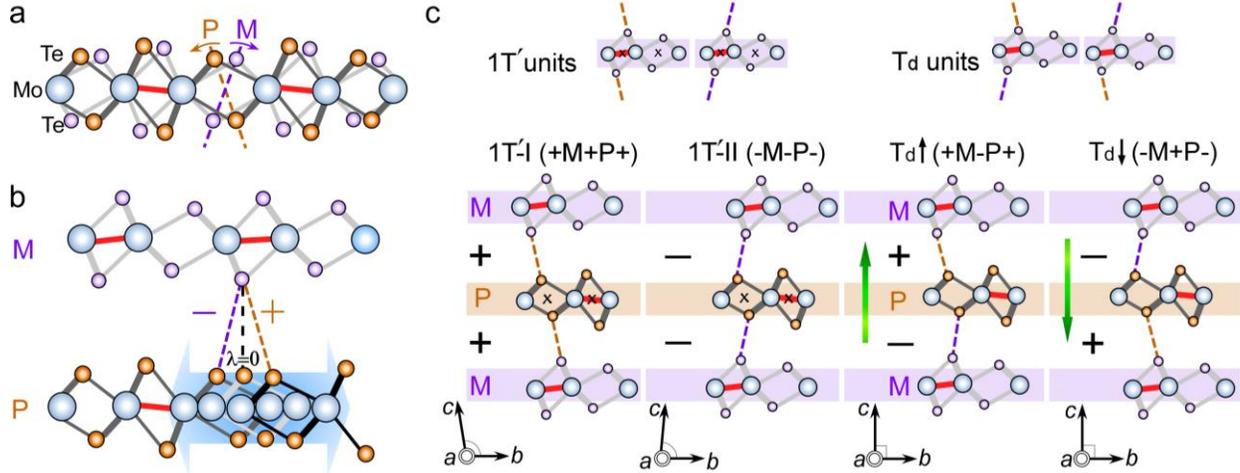

**Figure 1 (double column) Flexible layer-structured MoTe$_2$. a**, Schematic models of a single layer of MoTe$_2$ with either P (Plus, CCW rotation) or M (Minus, CW)-type Te octahedral deformations in the cross-section view. Mo, blue; Te of P-layer, orange; Te of M-layer, purple. The red lines indicate Mo-Mo zigzags along the *a* axis. Orange and purple arc arrows represent the directions of the Te octahedral deformations. **b**, Schematic models of bilayer MoTe$_2$ with P-M and P+M configurations, counting from the bottom P layer. Gliding of the bottom P layer results in + (positive)/– (negative) interlayer shifts, where the signs refer to the CCW/CW displacement of Te-Te bonding lines. A zero-interlayer shear ($\lambda = 0$) corresponds to a centrosymmetric orthorhombic reference structure T$_0$. **c**, Top, examples of 1T' and T$_d$ units of a M layer. A 1T' unit requires the same sign of interlayer shearing (++ or --) while those of a T$_d$ unit are different. Bottom, three layers can glide individually to give four configurations. (1) 1T'-I, +M+P+ with *b-c* angle of ~93.5° and (2) 1T'-II, -M-P- with *b-c* angle of ~86.5°;[56] (3) T$_d$↑, +M-P+ and (4) T$_d$↓, -M+P- with orthogonal unit cells. The polarization along the ±*c* axis (green arrows) denotes as T$_d$↑ and T$_d$↓. Note that lattice *a* and *b* of the 1T' structure are switched to match the zigzag direction as in the T$_d$ phase (*b* > *a*). The symbol *x* marks the inversion center.

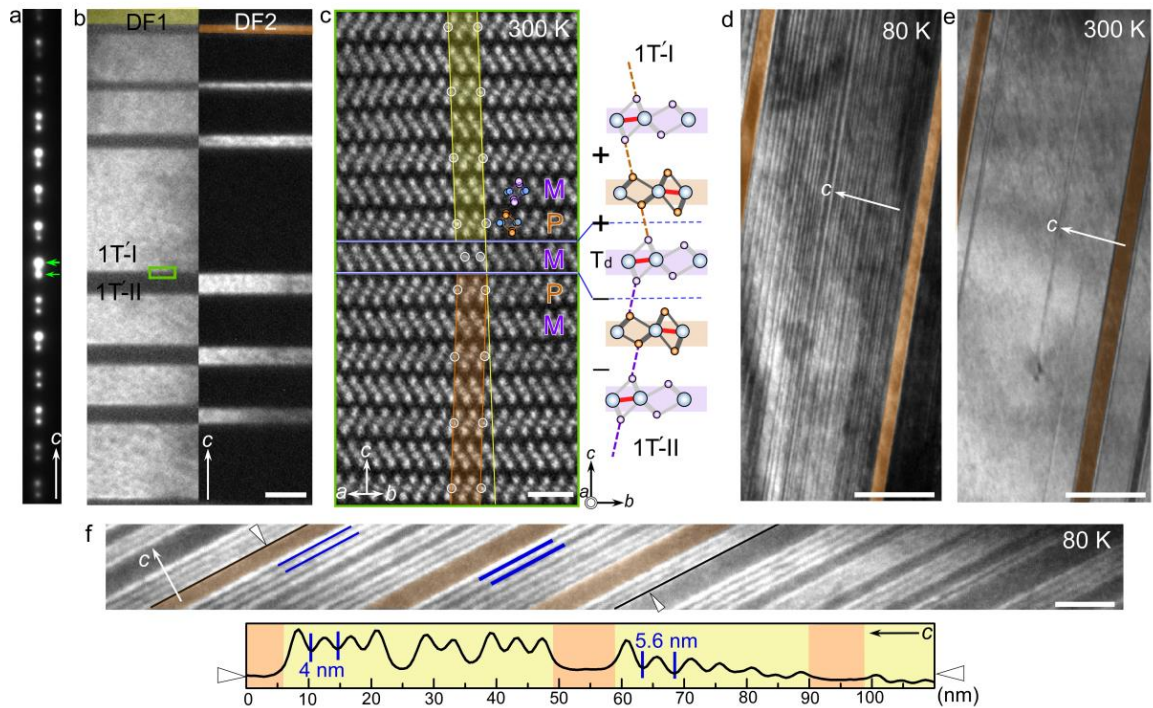

**Figure 2 (double columns) Phase domain walls along cross-section views at room and cryogenic temperatures. a**, A selected area electron diffraction (SAED) pattern of 1T' monoclinic twins, revealing spot splitting along the *c* axis. **b,** DF-images were taken using the strong and weak ($1\bar{1}2$) spots (green arrows) of variants 1T'-I and 1T'-II, denoted as DF1 and DF2, respectively. Yellow and orange false colors are added to aid the eye. The average twin width is on the order of a half-μm, and the two types of twin domains are typically unequal in size. Scale bar: 200 nm. **c**, The atomic-resolved HAADF-STEM image over one twin wall along the [110] zone axis. Overlaid color-coded 1T' unit cells defined by white-circled Mo atomic columns show a glide-reflection twin. The twin is composed of one M layer (marked by blue dashed lines) with anti-phase (+ -) interlayer shearing, connecting 1T'-I (yellow-shaded) and 1T'-II (orange-shaded) regions. The lattice model viewed along [100] is shown for clarity. Scale bar: 1 nm. **d, e**, DF-images taken at (**d**) 80 K and (**e**) 300 K. At 80 K, the phase-separated state is observed. The appearance of thin $T_d$ layers within the initial 1T' twins, revealing additional ($1\bar{1}2$) spots due to an orthogonal $T_d$ unit-cell as shown in Fig. 4a. Scale bar: 200 nm. **f**, A higher-magnification DF-image, showing superlattice-like $-(T_d)_m(1T')_n-$ (*m*, *n* = integer) nanoscale phase DWs at 80 K and the corresponding intensity profile between white arrow heads along the *c* direction, covering both 1T'-I (yellow) and 1T'-II (orange) twins. The image is rotated and enlarged to enhance the display. Scale bar: 30 nm.

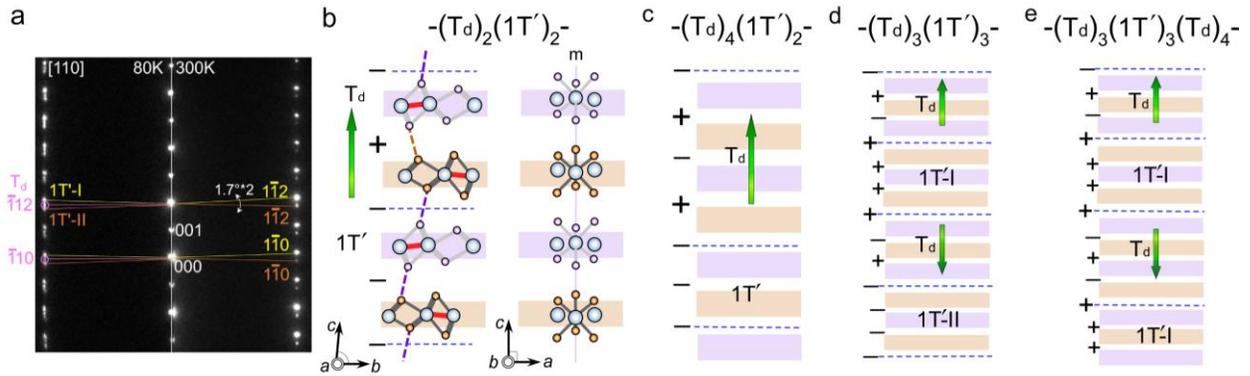

**Figure 3 (double columns) SAED pattern of MoTe$_2$ and the schematics of −(T$_d$)$_m$(1T')$_n$− 1T'/Td superlattice along *c* axis. a,** The presence of 1T'-I, 1T'-II twins and the T$_d$ phase are indicated with yellow, orange and pink straight lines. Though 1T' and T$_d$ shows the same extinction rules along [110], the orthogonal and non-orthogonal angles between ($1\bar{1}0$) and ($1\bar{1}2$) planes indicated by pink, yellow and orange lines unambiguously can be the fingerprint of T$_d$, 1T'-I and 1T'-II domains. **b,** The smallest −(T$_d$)$_2$(1T')$_2$− periodicity by symmetry. The configuration of +M-P+M+P+ corresponds to 4 layers, namely ~2.8 nm. Purple and orange blocks represent M and P layers, respectively. Blue dashed lines mark the phase domain walls. The mirror symmetry, perpendicular to the *a* axis, can be seen along [010] projection (unique axis *a*, when *b* > *a*). Experimentally, the thinnest periodicity is found to be 6 layers. **c,** Six layers can consist of either −(T$_d$)$_4$(1T')$_2$−, −(T$_d$)$_2$(1T')$_4$− or −(T$_d$)$_3$(1T')$_3$−. Considering the T$_d$ phase is the ground state at 80 K and the lattice mismatch as explained below. These considerations lead to the most likely domain assignment −(T$_d$)$_4$(1T')$_2$−. Polar T$_d$↑ can nucleate from either 1T'-I or 1T'-II without preference. The same rule is applied to T$_d$↓. **d,** Schematic model of the −(T$_d$)$_3$(1T')$_3$− periodicity. In order to maintain the three-layer periodicity, it requires the alternating of T$_d$↑, 1T'-I, T$_d$↓ and 1T'-II, which is unlikely to occur since the nucleation of 1T'-I (1T'-II) inside the existing 1T'-II (1T'-I) at low-temperature is un-favored. On the other hand, the simultaneous nucleation of T$_d$↑ and T$_d$↓ domains inside single 1T'-I domain is still possible as long as a change of periodicity occurs as shown in **e**. Experimentally, as shown in Fig. 2f, the periodicity does change within a single twin domain, which implies the possibility of the nucleation of opposite $T_d$ domains.

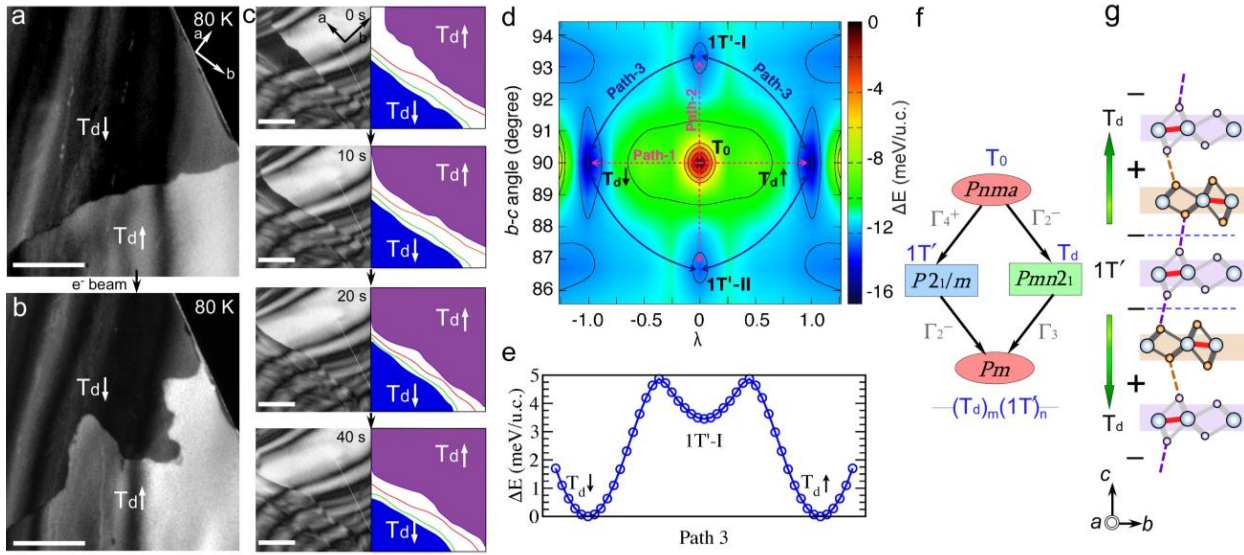

**Figure 4 (double columns) Polar domain and domain wall kinetics under e⁻ beam along plane views at 80 K. a**, DF-image of two $T_d$ domains with bright and dark contrasts. **b**, An immediate DF-TEM image of the same area after exposure to a focused e⁻ beam, showing DW motion accompanying the shrinkage of the dark-contrast $T_d\downarrow$ domain. **c**, Sequential snapshots obtained from the in-situ video, revealing representative polar DW motions after exposure to a focused **e⁻** beam. Four visible DWs are outlined in purple, red, green and blue, which represent the DWs between gradient domains of different $T_d\uparrow/T_d\downarrow$ volume fractions along the *c* axis. The $T_d\uparrow$ dominated (purple-shaded) domain is favored and expanded under electron beam while the $T_d\downarrow$ dominated (blue-shaded) domain has shrunk. Scale bar, 500 nm. **d**, The potential energy surface as a function of the normalized interlayer displacement ($\lambda$) and *b-c* cell angle. The color scale denotes energy with respect to the high-energy peak $T_0$ phase. **e**, The energy profile from the $T_d\downarrow$ to $T_d\uparrow$ transition along the lowest energy path-3. **f**, The family tree of the crystallographic group-subgroup relation. **g**, The schematic model of a $T_d\uparrow$ and $T_d\downarrow$ junction along the *c* axis, containing one 1T' unit as bridge. Blue dashed lines mark the phase boundaries.

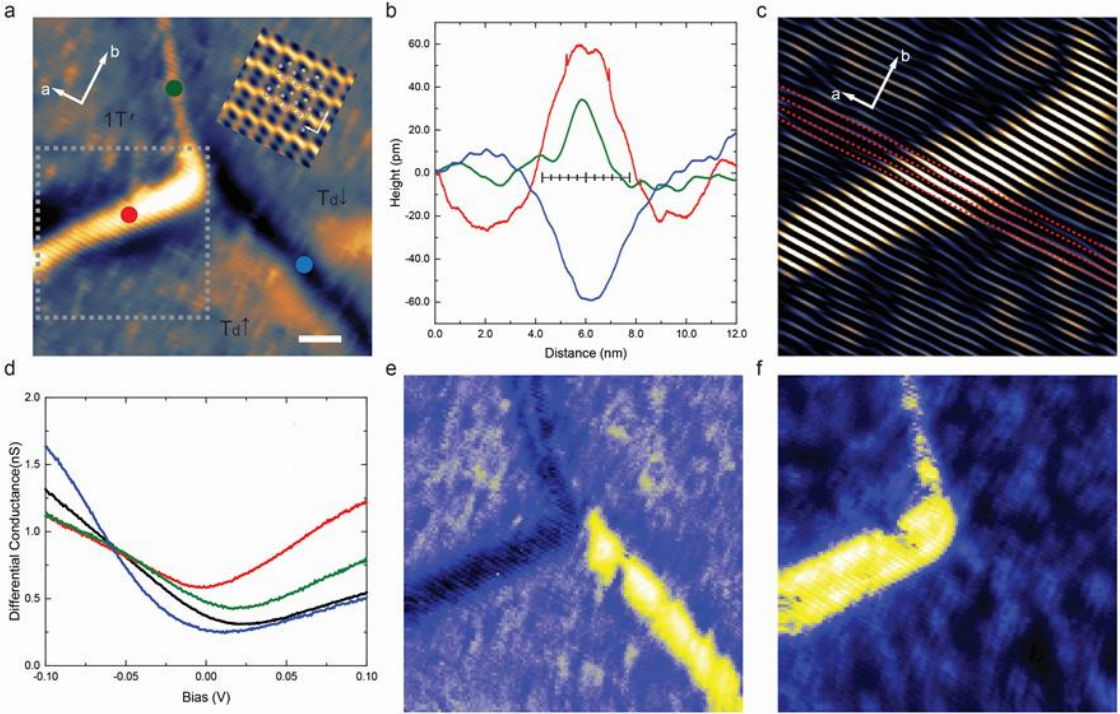

**Figure 5. (double columns) STM spectroscopic features of lateral phase domains and DWs in MoTe$_2$:Fe. a**, STM topography showing a junction of three domains and DWs. Inset: atomic resolution image of MoTe$_2$ showing Mo-Te-Mo chains. Scale bar: 5 nm. **b**, Height profile across each of three DWs obtained from red, green and blue dots in (**a**). Red and green represent the first-type protruded DWs and blue belongs to the second type. A ruler at the center shows the length of ten-unit cells for comparison. **c**, Fourier filtered topography of the first type DW from the dashed rectangle in (**a**). Red dashed lines mark ideal chain directions and the topography on the protruded area reveals the deviation of chain from the ideal straight line. **d,** Differential conductance obtained from each DW in (**a**). Red and green curves: the first-type protruded DWs; Blue curve: the second-type depressed DW; Black: the averaged curve obtained inside a domain, normalized at -100 mV, 100 pA. **e**,**f**, Spatial mapping of differential conductance at -100 mV (**e**) and +100 mV (**f**), normalized at -50 mV, 100 pA.

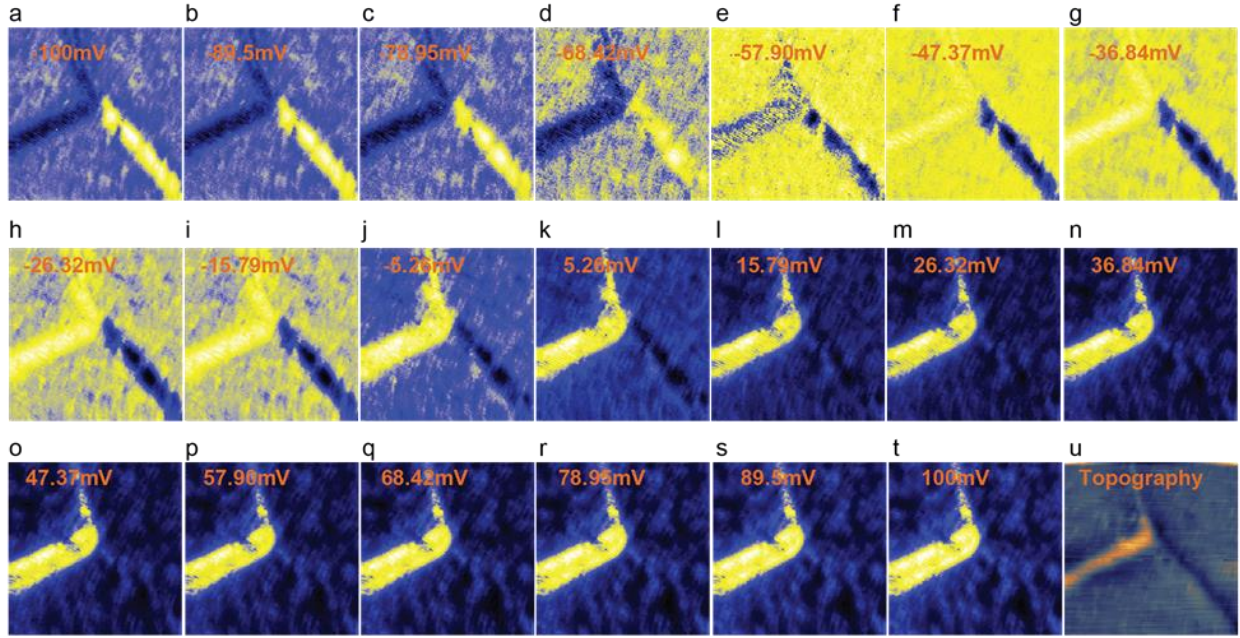

**Figure 6. (double columns) Spatial variation of local density of state near the junction. a-t,** A set of differential conductance maps obtained near a junction of three domains and DWs. All of spectra are normalized to the same tunneling resistance with -50 mV and 100 pA. Differential conductance spectra are obtained by demodulated lock-in signal using 10 mV of oscillation at 20 different biases from -100 to 100 mV. Twenty lock-in measurements are interlaced in between each line-scan of the topography shown in (**u**) to minimize the effect of thermal drift. There appear two different characteristics from the DWs. Two DWs on the upper left corner show depressed local density of state at energies below -50 mV, and start to show enhanced conductance above -50 mV. On the contrary, the DW in the right bottom corner shows the opposite feature, i.e. enhanced below -50 mV and suppressed above -50 mV. **u**, STM topography of the same area with the 40×40 nm$^2$ image size.

## Methods

Sample preparation

1T'-MoTe$_2$ single crystals were grown using the flux method. Well ground Mo (Alfa Aesar, 99.9 %) and Te (Alfa Aesar, 99.9 %) powders were mixed with sodium chloride (NaCl, Alfa Aesar, 99.9 %) in an alumina crucible, which was sealed in a quartz tube under vacuum. Crystallization was conducted from 1100 to 960 °C for 12 hrs, following a 0.5 °C/h cooling rate to 960 °C and then a rapid cooling to room temperature by placing the quartz tube in water (quenching).

Ribbon-like crystals (3*0.5*0.1 mm$^3$) with shiny surfaces were obtained. 1T'-MoTe$_2$:Fe single crystals were grown using a similar process with the starting composition of Fe$_{0.3}$MoTe$_2$, but tends to form ten times smaller in size and less cleavability. From the analysis of STM images, the resulting estimation of the real composition of Fe impurities is ~1.06 % (the details are given in Supplementary Fig. 2). The electrical transport measurements (along the *b* axis) were taken with the standard four-probe technique using Au paste as electrodes. Temperature was controlled by using a Physical Properties Measurement System (PPMS-9, Quantum Design), are consistent with the results in literature[32,33].

TEM measurements

Crystal structure, electron diffraction and domains were examined by transmission electron microscopy (TEM) in side view and plane view. Plane-view specimens were obtained by scotch-tape exfoliation, while side-view specimens were fabricated as follows. First, two silicon slabs and one MoTe$_2$ thin plate were clamped and glued together using epoxy bond (Allied, Inc) with sides facing each other to make a sandwich structure. The MoTe$_2$ sandwich was further thinned down by mechanical polishing, followed by Ar-ion milling, and studied using a JEOL-2010F field-emission TEM quipped with a low-*T* sample stage and a room-*T* double-tilt sample stage. We observed in-plane polar domains by DF-TEM imaging taking $g_{1\pm} = \pm(1, 2, \bar{1})$ spots along the [101] direction, ~14° tilting from the [001] zone and the side-view twin domains using $g_{2\pm} = \pm(1, \bar{1}, 2)$ spots along the [110] direction, ~60° tilting from the [100] zone. HAADF-STEM imaging with atomic-column resolution was carried out using the field-emission JEOL-2100F microscope equipped with a spherical aberration Cs corrector. All images are raw data. HAADF-STEM images were acquired in two conditions: 512×512 with 0.019 nm and 0.015 nm/pixel with collection angle between 80-210 mrad.

STM measurements

STM and spectroscopy measurements were performed at liquid nitrogen temperature using a Unisoku ultra-high vacuum SPM System (USM-1500) with a cleaving stage in the chamber. A Cu(111) sample that is cleaned by repeated cycles of sputtering and annealing prior to scanning has been used as a reference sample. A Pt/Ir tip is heated by electron beam bombardment in ultra-high vacuum condition to remove contaminations from air, and further treated on Cu(111) sample until it shows a metallic conductivity and the Cu(111) surface state spectroscopy. Fe doped MoTe$_2$ sample is fixed at a sample plate by silver epoxy (Epotek H20E) and a metal post is attached to the top with the same epoxy. Then the sample is introduced to ultra-high vacuum chamber and cleaved at room temperature in the cleaving stage followed by insertion to LN$_2$ cooled STM head. Differential conductance is measured by modulation of bias and demodulation of tunneling current using lock-in technique. (f = 611Hz, 10 mV with AC added to the bias).

Theoretical calculations

All the first-principles DFT calculations were performed using the Vienna ab initio simulation package (VASP) within the projected-augmented wave (PAW) framework[51,52]. We considered 6 valence electrons of Mo (4d$^5$5s$^1$) and 6 valence electrons of Te (5s$^2$5p$^4$) in the PAW pseudopotential. We used the PBEsol exchange-correlation functional to treat exchange and correlation effects[53]. A Monkhorst-Pack $k$ mesh of size 8×12×4 was used to sample the $k$-space, and 600 eV was used as the kinetic energy cutoff of the plane wave basis set. We also considered effects due to the on-site Coulomb interaction of Mo 4$d$ electrons, which were recently reported to be crucial in the precise determination of electronic structure of MoTe$_2$ near the Fermi-level. Within the DFT+U scheme, we used U = 2.4 eV and J = 0.4 eV to simulate Mo 4d electrons at the mean-field level. These values are reported to correctly describe the topological phase transitions and bulk electronic band structure of MoTe$_2$ near the Fermi-level[54]. The structures were optimized until the Hellmann-Feynman residual forces were less than 10$^{-4}$ eV/Å, and 10$^{-9}$ eV was defined as the convergence criterion for the electronic self-consistent calculations. Optimized lattice parameters and structural details are given in the Supplementary Table 1. Given the symmetry of 1T' and T$_d$ structures, $a$ and $b$ lattice vectors are interchangeable. In the DFT calculations, we used a convention in which Mo-Mo zigzag run along the $b$-lattice vector ($a > b$). The phonon calculations were performed using the finite-difference approach as implemented in the VASP software. Supercell of size 2×4×1 was used for phonon calculations, and PHONOPY code was used for the post-processing of phonons[55]. All the inner-coordinates of atoms were fully optimized, except for the modulated structures along the unstable phonon mode at Γ, which were optimized while keeping the coordinates of atoms frozen in the direction of modulation vectors. However, T$_d$↑ and T$_d$↓ structures were further optimized without any constraints.

**Acknowledgements**

The work at Rutgers was funded by the Gordon and Betty Moore Foundation's EPiQS Initiative through Grant GBMF4413 to the Rutgers Center for Emergent Materials and by NSF DMREF Grant No. DMR-1629059.

**Author contributions**

F.T.H. conducted the TEM experiments. S.J.L. carried out the STM observations. S.S., J.K., K.R. and D.V. carried out the theoretical analysis. J.W.K. performed transport measurements. L.Z. synthesized single crystals. F.T.H. and M.W.C performed the STEM observations. F.T.H., S.J.L., S.S., D.V. and S.W.C. wrote the manuscript. S.W.C. initiated and supervised the research.

**Data availability**

The authors declare that all source data supporting the findings of this study are available within the article and the Supplementary information file.

**Competing interests**

The authors declare no competing interests.

# Supplementary Information

# Polar and phase domain walls with conducting interfacial states in a Weyl semimetal MoTe$_2$

Huang *et al.*

**Supplementary Note 1 The origin of the uncompensated dipole in $T_d$-MoTe$_2$**

In this section, we explain the polar distortion in semimetallic MoTe$_2$ from the crystallographic point of view. The geometric structure of centrosymmetric 1T'-I MoTe$_2$ with + + interlayer pattern is illustrated in Supplementary Figure 1a. Each Mo ion (blue spheres) sits in a distorted Te octahedron. Because of the Mo-Mo metallic bonding, Mo ions shift off the center of the distorted Te octahedra. Green arrows indicate the $\pm z$ displacement of Mo ions away from the average $z$ positions of the neighboring Te ions. Since the top and bottom Te layers (yellow and red spheres) are confined by space inversion symmetry (the inversion center located at the Mo-Mo bond marked by x), the $\pm c$ dipole moment is cancelled out. The intrinsic difference between 1T' and $T_d$ arises from the interlayer gliding of the structure. As long as the interlayer gliding is arranged in either + - or - + pattern, it will break the inversion symmetry while any additional vertical displacement is not required as shown in Supplementary Figure 1b. The non-centrosymmetric structure is also clear when tracking the Te-Te interlayer bonding outlined by orange and purple dashed lines (Supplementary Figure 1b). The absence of any inversion center induced by the interlayer gliding correlated with the fact that $\pm c$ directions are geometrically non-equivalent. It does not necessarily accompany the existence of a net dipole moment, in the sense that assuming each P or M layer is as rigid and symmetric as that of 1T' (Supplementary Figure 1a inset). Thus, any interlayer gliding is not sufficient to explain the polar origin in the $T_d$ phase.

The inset of Supplementary Figure 1b demonstrates that the polar distortion is, in fact, caused by the local asymmetric bonding environment between the top Te (green and dark blue spheres) and bottom Te (yellow and red spheres) ions around the Mo ions. Because of the broken inversion symmetry, now the top and bottom Te are symmetry independent as indicated by different colors. Consequently, it gives additional vertical degree of freedom of those Te sites and leads to a net dipole moment along the $c$ axis. The estimation of the net dipole moment based on the $T_d$ structure [ref. 34 in the main text] is $3.6 \times 10^{11}$ e$^-$ cm$^{-2}$ (= 0.058 μC cm$^{-2}$). Note that a recent report on monolayer MoTe$_2$ with the $d$1T trimerized structure can achieve 0.68 μC cm$^{-2}$ from the DFT calculations [ref. 38 in the main text]. As discussed above, the asymmetric Te bonding environment of the $T_d$ phase is the reason that moves the average negative center away from the Mo-Mo center even though the magnitude is almost negligible compared with traditional ferroelectric perovskite (three orders smaller than BaTiO$_3$). Despite the fact that

MoTe$_2$ is semimetal, the charge distribution is expected to be highly anisotropic in the van der Waals layered structure. Therefore, the dipole-dipole interaction in the polarization direction may not be screened as a similar mechanism proposed in ferroelectric metal LiOsO$_3$ [ref. 23 in the main text]. Finally, we note that two distinct surfaces of MoTe$_2$ at low temperatures have been reported in literatures [refs. 29-30 in the main text], and they correspond to the T$_d$↑ and T$_d$↓ in this work.

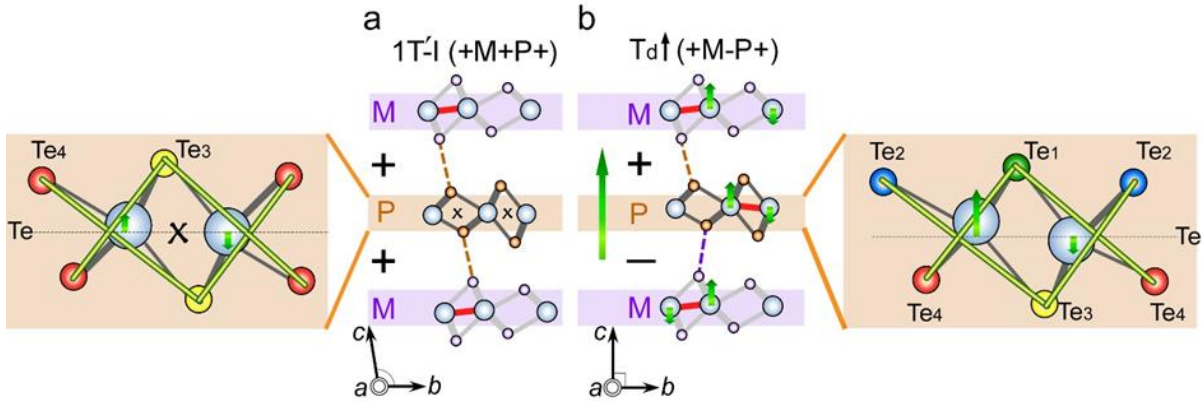

**Supplementary Figure 1. Schematics view of the macroscopic polarization in the T$_d$ phase. a**, 1T'-I and **b**, T$_d$↑ phase of MoTe$_2$. Mo, blue; Te of P-layer, orange; Te of M-layer, purple. The insets show the bonding environments of the P layers. A net out-of-plane dipole moment can be induced in the T$_d$ phase as a result of the slightly different Te environments on the top and bottom layers. Horizontal dotted lines shown in the insets indicate the average $z$ positions of Te octahedra. Green arrows correspond to the displacements of Mo ions along the $c$ axis. The symbol $x$ marks the inversion center.

**Supplementary Note 2: The persistence of phase coexistence in the presence of mechanical constraints**

As discussed in Figure 1 in the main text, the *b-c* angle in the monoclinic 1T' phase is associated with a mechanical glide of the layers. The macroscopic shear deformation during the phase transition involves the accumulated glide of the top layer relative to the bottom one by an amount which is proportional to the twin thickness along *c* axis. In the case of freely suspended or uncapped samples, the shear deformation can occur with no further energy cost, so that the entire sample can complete the structural transition into a single phase either of 1T' or $T_d$, depending on temperature. However, in a constrained situation such as a capped specimen, mechanical constraints will apply restoring forces that tend to resist the glide of the topmost layer from its initial position. For example, in this work, our cross-section TEM specimens were made by clamping two silicon slabs and one $MoTe_2$ thin plate together with sides facing each other. Experimentally, we always observe asymmetric 1T'-I and 1T'-II twin fractions (Figure 2b and Figure 2e in the main text) at room temperature, thus the 1T' to $T_d$ phase transition entails a shear deformation proportional to the thickness difference between the 1T'-I and 1T'-II domains. The resulting behavior depends on the strength of the elastic restoring forces. If these are strong enough that the net glide is almost forbidden, some fraction of 1T' phase should always remain even below $T_c$ as seen in this work. For instance, if the twin domain fractions are *x* and 1-*x* for the initial 1T'-I and 1T'-II phases respectively, with $x < ½$, then a fraction of 2*x* can transform to $T_d$ while leaving a fraction 1-2*x* of 1T'-II remaining. Therefore, the enhanced phase coexistence observed in the capped samples can be understood as an effect of mechanical constraints that tend to prevent the layer glide. In other words, we freeze in the coexistence of 1T' and $T_d$ phases in a quasi-periodic superlattice-like arrangement in the case of the capped specimen geometry at 80 K.

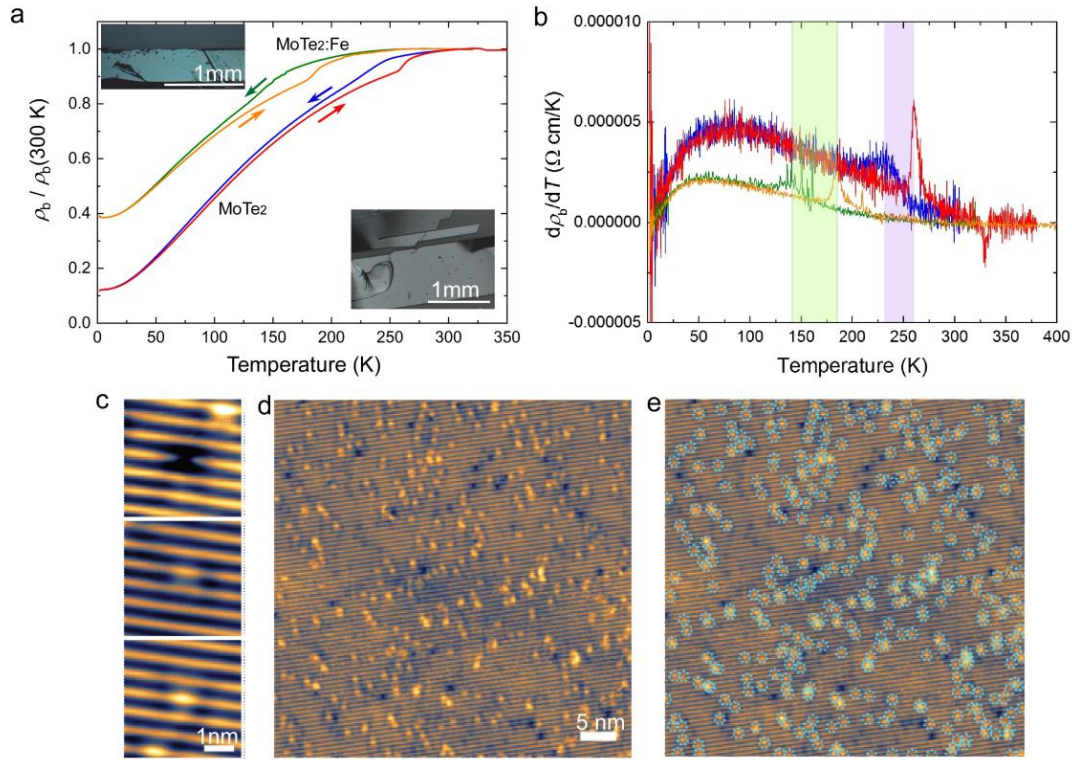

**Supplementary Figure 2. Chemically tunable WSM transition. a**, Temperature dependence of the *ab*-plane electrical resistivity of 1T'-$MoTe_2$ and 1T'-$MoTe_2$:Fe single crystals at ambient pressure. Measurements were carried out with the electric current fixed along the *b* crystallographic axis. Insets show photographic images of crystals. Scale bar, 1 mm. The 1T' to $T_d$ polar transition can be identified from the sudden decrease of resistance during cooling (blue and green curves) and the abrupt increase of resistance during warming (red and orange curves). Both resistivity curves show metallic behavior with a thermal hysteresis setting in below room T. **b**, Temperature derivative of resistivity, $d\rho(T)/dT$. Anomalous hysteresis loops are observed at 260-230 K in 1T'-$MoTe_2$ and 180-135 K in 1T'-$MoTe_2$:Fe single crystals. The phase transition can be tuned to lower temperature with Fe doping. **c,** Topographies of three typical defects found in 1T'-$MoTe_2$:Fe samples with scan parameters as -0.3V, 100 pA: top, a depression in the middle of a zigzag; middle, a protrusion connecting two zigzags; bottom, a protrusion in the middle of a zigzag. The first defect with the depression is identified as a Te vacancy while the bottom two defects with protrusion features are considered as a single Fe impurity as marked in (e). Scale bar: 1 nm. **d,e,** Estimation of Fe concentration is performed on a STM topography of 50 nm lateral size shown in (**d**) with scan parameters as -0.3V, 200 pA. The resulting estimation of the real composition of Fe impurities is ~1.06% (359 Fe atoms in a 50×50 $nm^2$ square area).

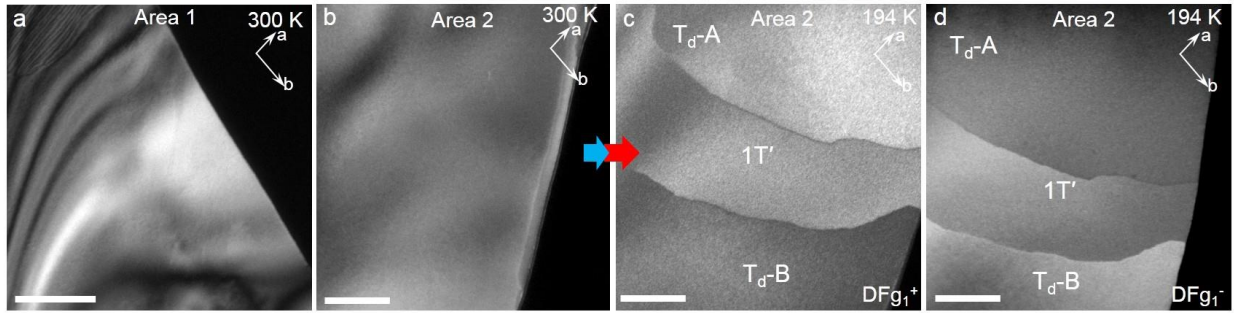

**Supplementary Figure 3. In-plane view of phase coexistence during a warming cycle. a**, DF-image showing no domain at room temperature in the nearby area of Figure 4a in the main text. Scale bar: 500 nm. **b**, DF-image showing no domain at room temperature in another area. During cooling down to 80 K and the following warming up process, the coexistence of 1T', $T_d\uparrow$ and $T_d\downarrow$ domains was found at 194 K. **b,c,** DF-image selecting $g_1^+ = (1, 2, \bar{1})$ spot in (**b**) and $g_1^- = (\bar{1}, \bar{2}, 1)$ spot in (**c**), showing the appearance of 1T' domain in the $T_d$ matrix during the warming process. The reversed contrasts of $T_d\uparrow$ and $T_d\downarrow$ are associated with a space-inversion breaking in the $T_d$ phase while the contrast of the central 1T' domain remains. Cycling the same TEM specimen through $T_c$ leads to a completely different polar domain patterns, indicating that the domain formation is not simply due to pinning by disorder such as chemical defects or dislocations. We note that 1T' shows up at 194 K which is lower than the data from the $T_c$ obtained in the transport measurement (shown in Supplementary Figure 2a) with a large coexistence temperature window of more than 65 K. On the other hand, a rather sharp transition has also been observed in a relatively thin specimen as shown in Supplementary Figure 4. Scale bar, 500 nm.

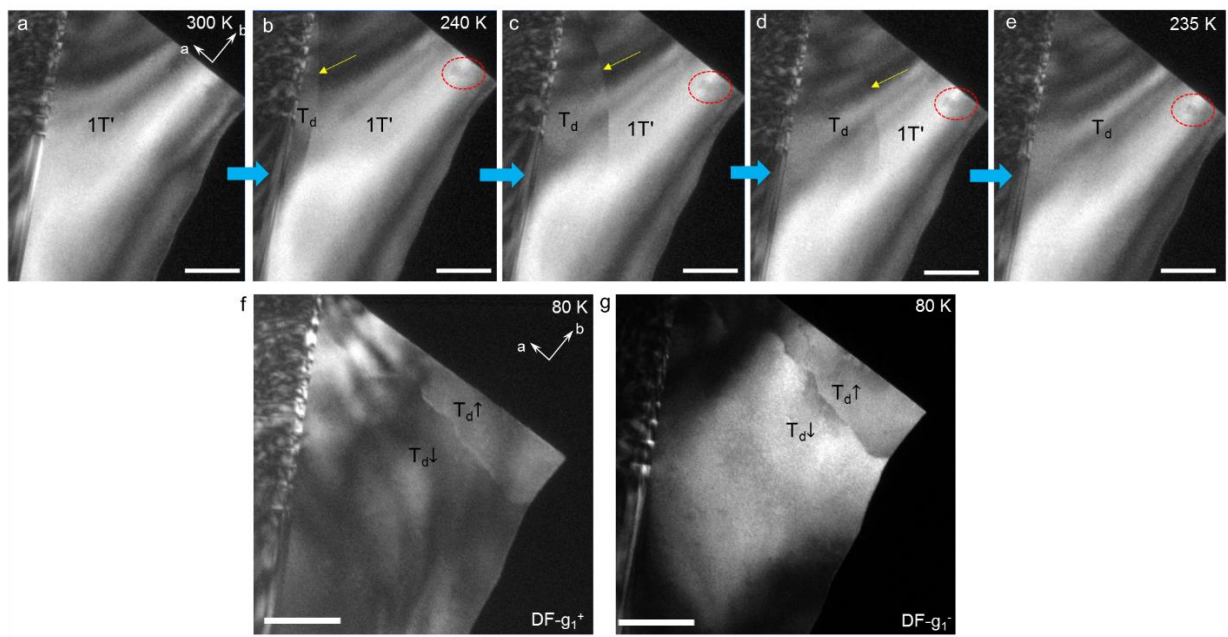

**Supplementary Figure 4 In-plane view of single 1T' to single T$_d$ domain evolution. a-e,** Sequential DF-images of in-situ cooling in moderately thin specimen (thickness < 30 nm), showing a rather sharp transition at ~260 K with a small temperature hysteresis (~5 K). It is consistent with the anisotropic phase evolution with a preferred layer-by-layer shearing path as discussed in the main text. A new T$_d$ domain nucleates and grows from the left to the right, leading to a bright to dark change in the domain contrast. The yellow arrow indicates the phase domain walls between the 1T' and T$_d$ phases. In term of the electron beam effect, if we start with 1T' phase, a beam-induced damaging instead of phase switching occurred as indicated by red circles. No phase transition is observed at room temperature. **f,g,** DF-images of beam induced T$_d$ domains after a beam exposure at the specimen edge. A new T$_d$ domain appears at the edge. Both T$_d$ domains show reversed contrasts when selecting g$_1^+$ spot (**f**) and g$_1^-$ spot (**g**) for imaging at 80 K. We found that the e$^-$ beam induced domains only occur in the T$_d$ phase and not in the 1T' phase. The nucleated polar DWs tend to be parallel to the *a* axis. Note that the assignments of T$_d$↑ and T$_d$↓ is only for simplicity and needs further confirmation by measurements such as atomic-resolved STEM imaging. Scale bar, 500 nm.

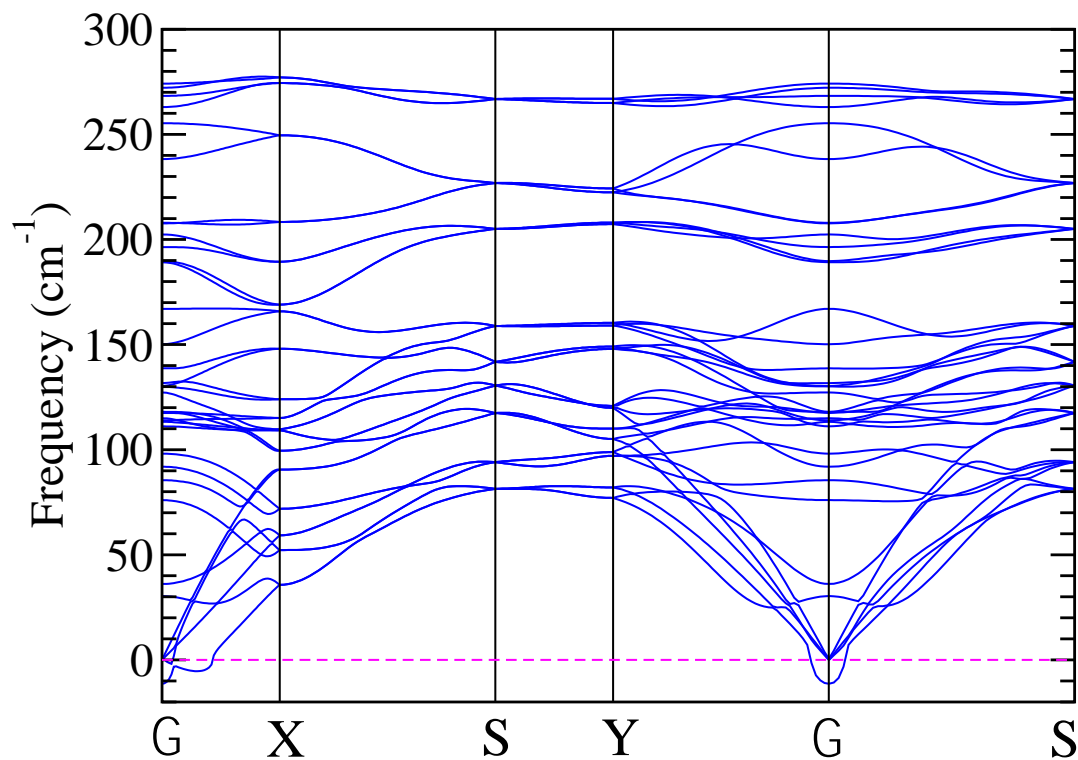

**Supplementary Figure 5 Phonon frequencies of the $T_0$ phase calculated along the high-symmetry directions of momentum space.** Two instabilities can be observed in the phonon band structure: (1) An unstable polar phonon mode present at the Γ point causing interlayer displacement of alternating MoTe$_2$ layers, and (2) an unstable acoustic mode along the Γ-X direction indicating the elastic instability present in the $T_0$ phase.

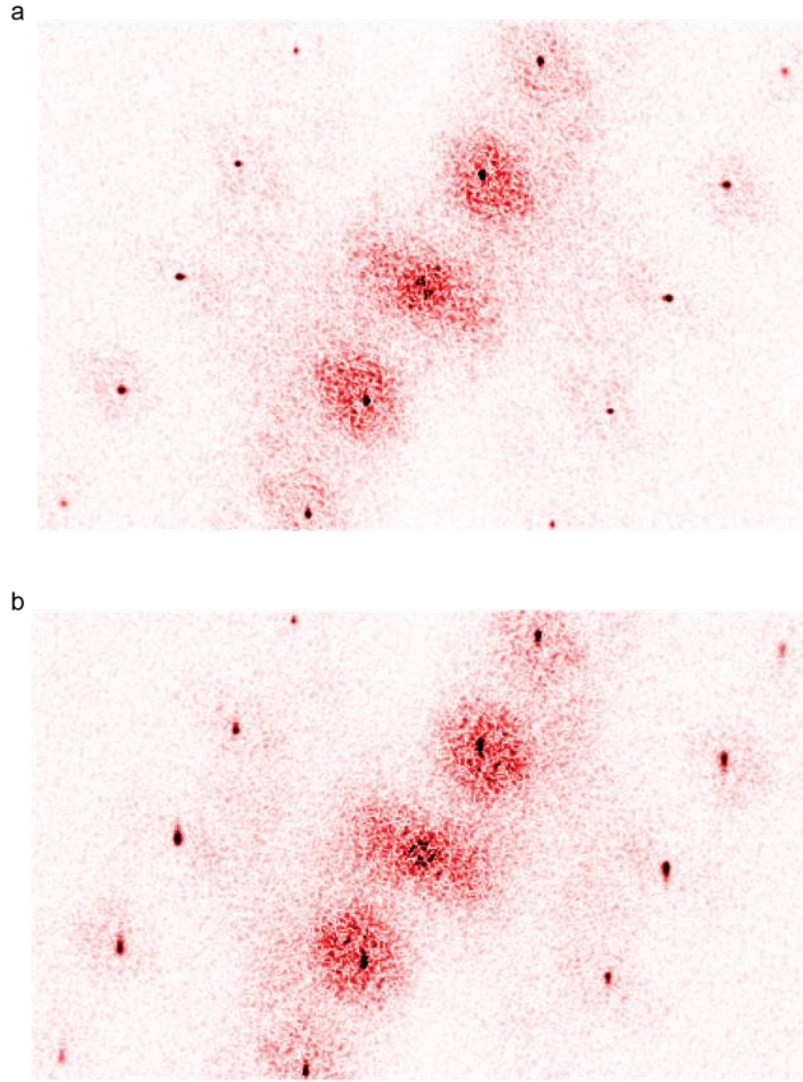

**Supplementary Figure 6 Quasiparticle interference patterns at 77 K.** We have tried to get a meaningful difference of quasiparticle interference pattern from multiple domains, however, all measurements showed similar results within our system's resolution or at the measurement temperature. **a,b,** Two representative patterns obtained from two nearby domains. There are clear 2×1 superlattice peaks appearing as a rectangular pattern of black dots. Although long-range featured near the center point differ in detail from different domains, we attribute the change to the effect of different atomic defects such as Fe dopants. The observed pattern near the center is more diffuse than the reported patterns of $MoTe_2$ [ref. 29 in the main text], which supports the effect of dopants in our measurement.

**Supplementary Table 1:** DFT+U optimized crystal parameters.

| MoTe$_2$ | Space group | Lattice parameters (Å) | Cell angles (degrees) |
|---|---|---|---|
| T$_d$↑/T$_d$↓ | 31 | $a = 6.329, b = 3.450, c = 13.485$ | $\alpha = 90°, \beta = 90°, \gamma = 90°$ |
| 1T'-I | 11 | $a = 6.329, b = 3.450, c = 13.485$ | $\alpha = 90°, \beta = 86.4, \gamma = 90°$ |
| 1T'-II | 11 | $a = 6.329, b = 3.450, c = 13.485$ | $\alpha = 90°, \beta = 93.6, \gamma = 90°$ |
| T$_0$ | 62 | $a = 6.329, b = 3.450, c = 13.485$ | $\alpha = 90°, \beta = 90°, \gamma = 90°$ |

**Direct atomic coordinates:**

**T$_0$ phase**

|    | x | y | z |
|---|---|---|---|
| Mo | 0.11349 | 0.25000 | 0.00690 |
| Mo | 0.25103 | 0.75000 | 0.50696 |
| Mo | 0.75096 | 0.75000 | 0.99309 |
| Mo | 0.61343 | 0.25000 | 0.49305 |
| Te | 0.50373 | 0.25000 | 0.10509 |
| Te | 0.00793 | 0.75000 | 0.15199 |
| Te | 0.50704 | 0.75000 | 0.34798 |
| Te | 0.00286 | 0.25000 | 0.39494 |
| Te | 0.36073 | 0.75000 | 0.89490 |
| Te | 0.85653 | 0.25000 | 0.84799 |
| Te | 0.35740 | 0.25000 | 0.65202 |
| Te | 0.86158 | 0.75000 | 0.60507 |

**T$_d$↑(T$_d$↓) phase**

|    | x | y | z |
|---|---|---|---|
| Mo | 0.14865 (0.07433) | 0.25000 (0.25000) | 0.00709 (0.00641) |
| Mo | 0.21580 (0.29012) | 0.75000 (0.75000) | 0.50714 (0.50645) |
| Mo | 0.79010 (0.71579) | 0.75000 (0.75000) | 0.99355 (0.99286) |
| Mo | 0.57435 (0.64866) | 0.25000 (0.25000) | 0.49360 (0.49291) |
| Te | 0.54017 (0.46562) | 0.25000 (0.25000) | 0.10525 (0.10506) |
| Te | 0.04564 (0.97440) | 0.75000 (0.75000) | 0.15258 (0.15174) |
| Te | 0.47442 (0.54566) | 0.75000 (0.75000) | 0.34826 (0.34742) |
| Te | 0.96563 (0.04018) | 0.25000 (0.25000) | 0.39494 (0.39475) |
| Te | 0.39882 (0.32427) | 0.75000 (0.75000) | 0.89490 (0.89471) |
| Te | 0.89001 (0.81881) | 0.25000 (0.25000) | 0.84821 (0.84737) |
| Te | 0.31883 (0.39003) | 0.25000 (0.25000) | 0.65263 (0.65179) |
| Te | 0.82429 (0.89884) | 0.75000 (0.75000) | 0.60529 (0.60510) |

**1T'-I (1T'-II) phase**

|    | x | y | z |
|----|---|---|---|
| Mo | 0.11221  (0.11066) | 0.25000  (0.25000) | 0.00726  (0.00631) |
| Mo | 0.25428  (0.25180) | 0.75000  (0.75000) | 0.50628  (0.50723) |
| Mo | 0.75177  (0.75429) | 0.75000  (0.75000) | 0.99276  (0.99370) |
| Mo | 0.61063  (0.61215) | 0.25000  (0.25000) | 0.49370  (0.49276) |
| Te | 0.51773  (0.48836) | 0.25000  (0.25000) | 0.10500  (0.10532) |
| Te | 0.02787  (-0.00714) | 0.75000  (0.75000) | 0.15274  (0.15235) |
| Te | 0.49236  (0.52739) | 0.75000  (0.75000) | 0.34765  (0.34728) |
| Te | -0.01207  (0.01729) | 0.25000  (0.25000) | 0.39464  (0.39502) |
| Te | 0.34624  (0.37660) | 0.75000  (0.75000) | 0.89500  (0.89468) |
| Te | 0.83615  (0.87208) | 0.25000  (0.25000) | 0.84728  (0.84767) |
| Te | 0.37254  (0.33659) | 0.25000  (0.25000) | 0.65233  (0.65271) |
| Te | 0.87700  (0.84664) | 0.75000  (0.75000) | 0.60535  (0.60498) |